\documentclass[sn-mathphys,pdflatex]{sn-jnl}% Math and Physical Sciences Reference Style
\usepackage{physics}
\jyear{2021}%

\begin{document}

\title{Dichroism in time-resolved ARPES and valence band orbital nature in $\rm{BaNiS_2}$}


\author*[1]{\fnm{Jiuxiang} \sur{Zhang}}\email{jiuxiang.zhang@universite-paris-saclay.fr}

\author[1]{\fnm{Zhesheng} \sur{Chen}}

\author[1]{\fnm{Jonathan} \sur{Caillaux}}

\author[2]{\fnm{Yannick} \sur{Klein}}

\author[2]{\fnm{Andrea} \sur{Gauzzi}}

\author[3]{\fnm{Azzedine} \sur{Bendounan}}

\author[3]{\fnm{Amina} \sur{Taleb-Ibrahimi}}

\author[4]{\fnm{Luca} \sur{Perfetti}}

\author[1]{\fnm{Evangelos} \sur{Papalazarou}}

\author*[1]{\fnm{Marino} \sur{Marsi}}\email{marino.marsi@universite-paris-saclay.fr}

\affil*[1]{\orgname{Université Paris-Saclay, CNRS, Laboratoire de Physique des Solides}, \orgaddress{\postcode{91405}, \state{Orsay}, \country{France}}}

\affil[2]{\orgname{Sorbonne Université, CNRS, IMPMC, IRD, MNHN}, \orgaddress{\postcode{75252}, \state{Paris}, \country{France}}}

\affil[3]{\orgname{Synchrotron SOLEIL}, \orgaddress{\postcode{91192}, \state{Gif-sur-Yvette}, \country{France}}}

\affil[4]{\orgname{Ecole Polytechnique, Institut Polytechnique de Paris,  CEA/DRF/lRAMIS, CNRS, Laboratoire des Solides Irradiés}, \orgaddress{\postcode{91128}, \state{Palaiseau}, \country{France}}}

\abstract{Time-resolved ARPES gives access to the band structure and ultrafast dynamics of excited electronic states in solids. The orbital character of the bands close to the Fermi level is essential to understand the origin of several exotic phenomena in quantum materials. By performing polarization dependent time- and angle-resolved photoemission spectroscopy and by analizing the chirality of the photoelectron yield for two different crystal orientations, we identify the orbital character of bands below and above the chemical potential for the Dirac semimetal $\rm{BaNiS_2}$. Our results illustrate how the control and understanding of matrix elements effects in time-resolved photoemission spectroscopy can be a powerful tool for the study of quantum materials.}




\maketitle

\section{Introduction}\label{sec1}

\begin{figure*}[h]%
	\centering
	\includegraphics[width=0.9\textwidth]{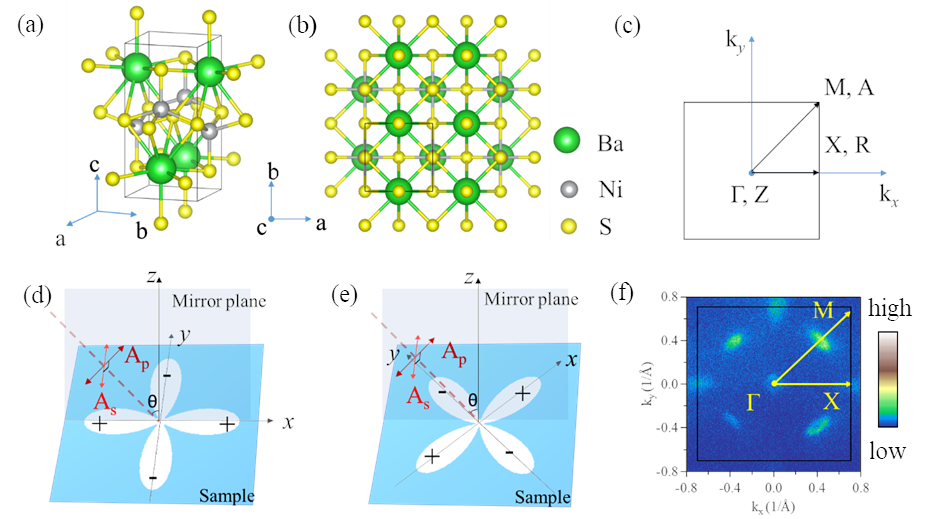}
	\caption{(a) 3-dimensional view of the unit cell. (b) Top view of 2×2 unit cells. (c) (2D) projection of the Brillouin zone in the $k_x$, $k_y$ plane. (d) Illustration of the spatial symmetry of $d_{x^2-y^2}$ with respect to the mirror plane in $\rm{\Gamma X}$ configuration. (e) The same but in $\rm{\Gamma M}$ configuration.  (f) Fermi surface of $\rm{BaNiS_2}$ acquired with 70 eV photon energy}\label{fig1}
\end{figure*}

$\rm{BaNiS_2}$ has recently generated increasing attention because of its peculiar properties related to the combined effects of crystal field and strong spin-orbit coupling \cite{bib1,bib2,bib3}, resulting in hidden Rashba-split spin-polarized bands. Furthermore, the presence of tunable quasi two-dimensional Dirac cones stemming from a nonsymmorphic symmetry of the lattice has been recently demonstrated \cite{bib4}. It has been shown that the tunability of Dirac cones origins from the charge transfer generated by the hybridization between $p$ and $d$ orbitals \cite{bib5}, which makes $\rm{BaNiS_2}$ a promising system to functionalize Dirac stats by manipulating the strength of electron correlations. Furthermore, $\rm{BaNiS_2}$ is the metallic precursor of $\rm{BaCo_{1-\textit{x}}Ni_\textit{x}S_2}$, which presents great similarities with high-Tc cuprates, but no superconductivity \cite{bib6,bib7,bib8}. Understanding electron-electron correlations will help obtain further insight about all these peculiar properties. In Fig. \ref{fig1}(a)(b), we show the crystal structure of $\rm{BaNiS_2}$ \cite{bib9} and in Fig. \ref{fig1}(c) the two-dimensional projection of its Brillouin zone. 

The advent of time and angle-resolved photoemission spectroscopy (time-resolved ARPES) makes it possible to explore empty electronic states and their ultrafast dynamics  \cite{bib10}, extending to the study of excited states the distinctive features of ARPES. This technique provides a direct way towards the studies of band structure of crystalline solids, and to the effects of electron-electron correlations \cite{bib11}. In photoelectron spectroscopy, electrons in solids can be excited into the vacuum when they absorb photons with energy larger than the work function – an extension of the photoelectric effect. A spectrum measured with this technique contains information on the momentum and kinetic energy of photoelectrons; the intensity of the yield is proportional to the overlap integral between an initial state wave function $\Psi_i$ transformed according to the interaction operator $H_{int}$ and the final state wave function $\Psi_f$, and depends on the geometry of the experimental setup, the photon energy and the polarization of light \cite{bib11,bib12,bib13}, which can be written as following:

\begin{equation}
Intensity \propto \lvert M_{fi}^{\boldsymbol{k}} \rvert^2
\end{equation}

\begin{equation}
M_{fi}^{\boldsymbol{k}}=\bra{\Psi_f^{\boldsymbol{k}}} \boldsymbol H_{int} \ket{\Psi_i^{\boldsymbol{k}}} \propto \bra{\Psi_f^{\boldsymbol{k}}} \boldsymbol{A}\cdot\boldsymbol{p} \ket{\Psi_i^{\boldsymbol{k}}}
\end{equation}
where $\boldsymbol{A}$ is the electromagnetic vector potential, and $\boldsymbol{p}$ is the electronic momentum operator.
 
Having a better understanding of the relation between photoemission and spectral functions is crucial to extract more information on the electronic states, beyond the band structure $E = E(k)$, such as for instance band degeneracy and correlations. Furthermore, the photoelectron yield presents additional modulations in its intensity coming from matrix element effects  \cite{bib14}. The selection rules stemming from these effects can be actually exploited to obtain information on the orbital nature of the initial state by controlling some parameters used in the measurement \cite{bib15,bib16,bib17,bibOrbVO,bibSpinOrb}, either suppressing or emphasizing the photoelectron yield from specific electronic states. 

Selection rules state that since the final state is an even function with respect to the mirror plane (we assume that the photoelectron energy analyzer is in the mirror plane), orbitals that are odd (even) with respect to the mirror plane can be detected only with light polarization that has the same parity with respect to that plane. Here we use $d_{x^2-y^2}$ as example. Fig. \ref{fig1}(d)(e) shows the experimental geometry. In our geometry, a photon with p polarization has an incoming electric field that lies within mirror plane, while an $s$ polarized photon has an electric field perpendicular to the mirror plane, i.e., parallel to the sample surface: as a result, p polarization is even while s polarization is odd with respect to the mirror plane. A $d_{x^2-y^2}$ orbital has even parity when its ${xz}$ plane coincides with the mirror plane, therefore it can be detected with p polarization light. The overall matrix elements for d orbitals when the mirror plane is aligned with two high symmetry directions of the crystal are summarized in Table \ref{tab1}. By using state of the art electron analyzers, e.g., Angle Resolved Time-of-Flight (ARTOF) analyzers, one can simultaneously measure $E$, $k_x$ and $k_y$ without rotating the sample, making it possible to discriminate between matrix element and experimental geometry effects. By combining the rotation of the crystal and the control of light polarization, one can get a clear and unambiguous relation between the modulation of the photoemission intensity and the orbital characters. 

In this work, we use polarization dependent pump-probe ARPES to study the orbital character of the filled and excited electronic states of the Dirac semimetal $\rm{BaNiS_2}$. This information is obtained thanks to detailed measurements with well-chosen crystal orientations and different light polarizations and by comparing the results with theoretical electronic band calculations. 

\begin{table}[h]
	\begin{center}
		\begin{minipage}{\textwidth}
			\caption{Possibility of detecting 3d orbitals along two high symmetry directions with p and s polarization light}\label{tab1}
			\begin{tabular*}{\textwidth}{@{\extracolsep{\fill}}ccccccc@{\extracolsep{\fill}}}
				\toprule%
\multirow{2}{0.15\textwidth}{High symmetry direction} & \multirow{2}{*}{Light polarization} & \multicolumn{5}{c}{3d orbitals} \\ \cline{3-7} 
&   & $d_{x^2-y^2}$ & $d_{z^2}$ & $d_{xz}$ & $d_{yz}$ & $d_{xy}$ \\ \hline
\multirow{2}{0.15\textwidth}{$\rm{\Gamma M}$} &        p           &  No  &  Yes  &  Yes  & Yes  & Yes  \\
&         s          &  Yes &  No &  Yes  & Yes  &  No \\
\multirow{2}{0.15\textwidth}{$\rm{\Gamma X}$} &        p           &  Yes  & Yes   & Yes   & No  & No  \\
&         s          &  No  &  No  & No   & Yes  &  Yes \\ 
				\botrule
			\end{tabular*}
		\end{minipage}
	\end{center}
\end{table} 

\section{Experimental methods}\label{sec2}

High quality single crystals of $\rm{BaNiS_2}$ were grown by a self-flux method  \cite{bib18}, and the fresh surfaces were obtained by cleaving \textit{in-situ} along the (001) plane under ultrahigh vacuum, at a base pressure of $2 \times 10^{-10}$ mbar. The time-resolved ARPES experiments were performed at 85 K using a Scienta Omicron ARTOF-2 analyzer and a femtosecond Ti:Sapphire laser system FemtoSource XL 300. The near infrared pulses (1.55 eV) were upconverted to 6.2 eV, an energy sufficient to excite photoelectrons \cite{bibAPL}, and cascade frequency mixing in $\rm{\beta-BaB_2O_4}$ non-linear crystals and optics \cite{bibApplOpt}. Complementary measurements have been performed on the FemtoARPES setup and on the TEMPO beamline at the SOLEIL synchrotron.

The experimental geometry for the ARPES measurements is shown in Fig. \ref{fig1}(d)(e). For sake of simplicity, when the sample’s $\rm{\Gamma X}$ ($\rm{\Gamma M}$) direction is in the mirror plane, we call it $\rm{\Gamma X}$ ($\rm{\Gamma M}$) configuration, respectively. Various light polarization configurations were used during the measurements. Besides linearly polarized s and p photons (Fig. \ref{fig1}(d)(e)), we also used left and right circularly polarized photons generated by using a quarter-wave plate.

\section{Results and discussion}\label{sec3}

The data presented in Fig. \ref{fig1}(f) show the Fermi surface of $\rm{BaNiS_2}$ with reference to the indicated high symmetry directions in the a-b plane. It is possible to see the Fermi surface over the whole BZ, featuring four Dirac nodes along $\rm{\Gamma M}$, four oval pockets at X and one pocket at the $\rm{\Gamma}$ point. The data were obtained with synchrotron radiation at 70 eV photon energy and at 80 K.

As underlined by recent studies \cite{bib4,bib5,bib6}, the most important electronic properties of $\rm{BaNiS_2}$ are related to the Dirac cones and to the electron pocket at $\rm{\Gamma}$. We consequently focused on these bands when exploring in detail the polarization dependence of the whole angular photoemission yield, by using the ARTOF photoelectron detector of our ultrafast time-resolved ARPES setup.

\begin{figure}[h]%
	\centering
	\includegraphics[width=0.9\textwidth]{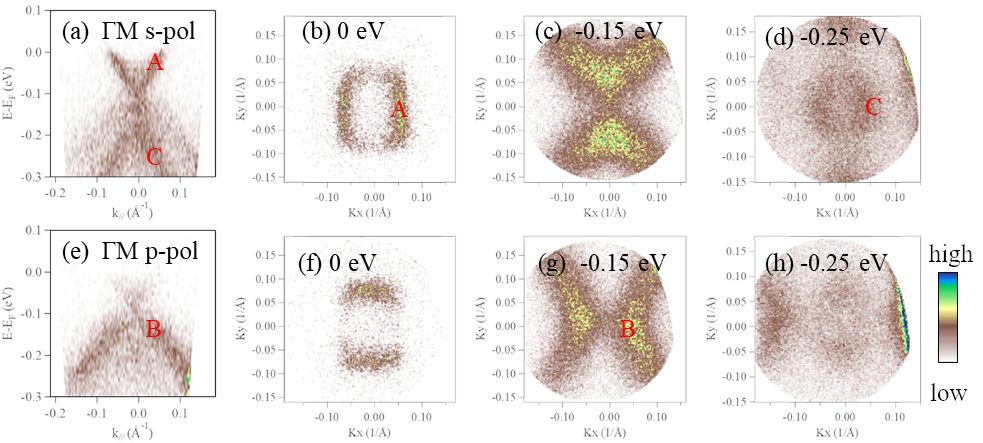}
	\caption{ARPES in $\rm{\Gamma M}$ configuration. (a) $E-k$ cut with s polarization. (b)-(d) CECs with s polarization at $E-E_F = 0$ eV, -0.15 eV and -0.25 eV, respectively. (e) $E-k$ cut with p polarization. (f)-(h) CECs with p polarization at $E-E_F = 0$ eV, -0.15 eV and -0.25 eV, respectively. }\label{fig2}
\end{figure}

\begin{figure}[htb]%
	\centering
	\includegraphics[width=0.9\textwidth]{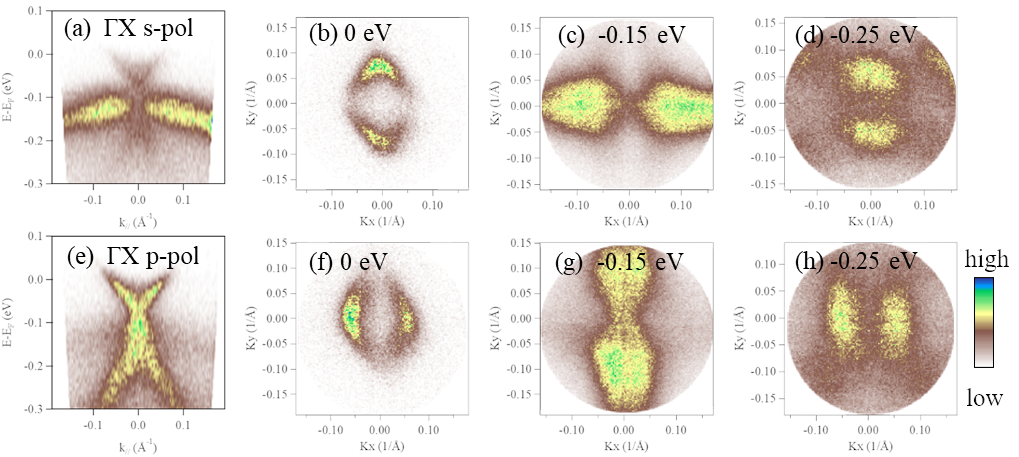}
	\caption{ARPES in $\rm{\Gamma X}$ configuration. (a) $E-k$ cut with s polarization. (b)-(d) CECs with s polarization at $E-E_F = 0$ eV, -0.15 eV and -0.25 eV, respectively. (e) $E-k$ cut with p polarization. (f)-(h) CECs with p polarization at $E-E_F = 0$ eV, -0.15 eV and -0.25 eV, respectively. }\label{fig3}
\end{figure}

Fig. \ref{fig2} (Fig. \ref{fig3}) shows photoemission data taken around $\rm{\Gamma}$ in $\rm{\Gamma M}$ ($\rm{\Gamma X}$) configuration, respectively. The data taken with s polarization are presented in the upper panels, while lower panels show the corresponding spectra taken in p polarization, respectively. Three bands can be resolved in Fig. \ref{fig2} and Fig. \ref{fig3}: one electron-like band A, two hole-like bands B and C. All three bands undergo very strong intensity variations when the light polarization is changed.

From the $E-k$ plots (Figs. \ref{fig2}(a) and \ref{fig3}(a) we can see that both bands A and C appear intense in s polarization in $\rm{\Gamma M}$ configuration and in p polarization in $\rm{\Gamma X}$ configuration, while they cannot be observed in p polarization in $\rm{\Gamma M}$ configuration and in s polarization in $\rm{\Gamma X}$ configuration. This means that the orbital character in these two bands is even with respect to mirror plane in $\rm{\Gamma X}$ configuration and odd with respect to mirror plane in $\rm{\Gamma M}$ configuration, which point to the $d_{x^2-y^2}$ orbital character based on Table \ref{tab1}. This result is consistent with the study of Santos-Cottin et al. \cite{bib1}. Note that both band A and band C have square shapes in constant energy contours (CECs) at selected energies $\rm {E-E_F = 0}$ eV and $E-E_F = -0.25$ eV, as shown in Fig. \ref{fig2}(b,f) and Fig. \ref{fig3}(b,f), but the orientation of two bands has 45-degree difference. The band B shows opposite behaviors, as it can be observed in p polarization in $\rm{\Gamma M}$ configuration and in s polarization in $\rm{\Gamma X}$ configuration (Figs. \ref{fig2}(a) and \ref{fig3}(a)). Therefore, the orbital character is even with respect to mirror plane in $\rm{\Gamma M}$ configuration and odd with respect to mirror plane in $\rm{\Gamma X}$ configuration, which could be attributed to $d_{xy}$ orbital character according to Table \ref{tab1}. 

\begin{figure}[htb]%
	\centering
	\includegraphics[width=0.9\textwidth]{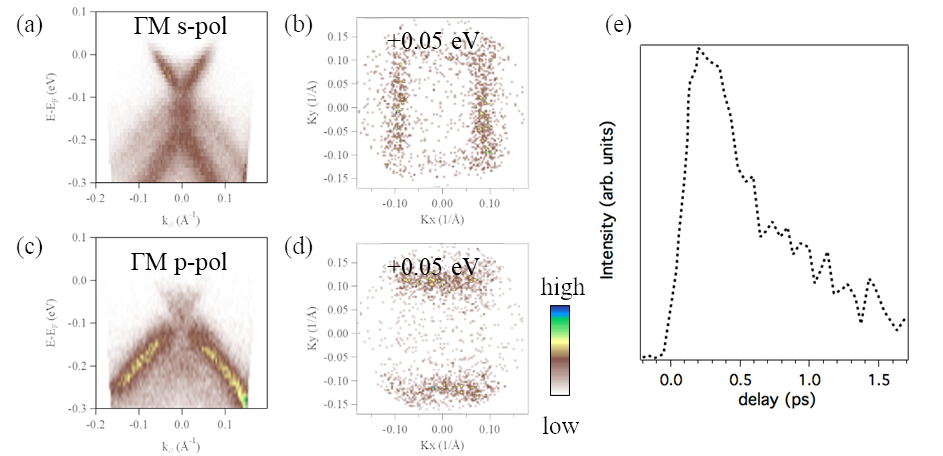}
	\caption{Pump-probe ARPES along $\rm{\Gamma M}$ at a delay of 250 fs (a) $E-k$ cut with s polarization. (b) CECs with s polarization at $E-E_F = +0.05$ eV. (c) $E-k$ cut with p polarization. (d) CECs with p polarization at $E-E_F = +0.05$ eV. (e) Time evolution of the intensity of the photoelectron yield above the chemical potential }\label{fig4}
\end{figure}

It should be emphasized that these conclusions can be extended to the empty electronic states that we have been able to access by optically pumping the system during the time-resolved ARPES measurements. In Fig. \ref{fig4} we present pump-probe ARPES data at a delay where we have maximum counts above Fermi level (Fig. \ref{fig4}(e)) and the relative linear dichroism contrast for CEC at $E-E_F = +0.05$ eV.

From ARPES data we could see that the electron pocket A and hole pocket C have square shapes in CECs, and from the crystal structure we could know that Ni atoms are surrounded by two pyramids, one formed by S atoms, and the other formed by Ba atoms (see Fig. \ref{fig1}). These two pyramids have a 45-degree difference in real space, and interestingly, the two bands A and C point at two directions with 45-degree difference. This naturally indicates a correlation between the directions of d-bands A and C, essentially, and the position of the S and Ba atoms with respect to the Ni atom.

\begin{figure}[htb]%
	\centering
	\includegraphics[width=0.6\textwidth]{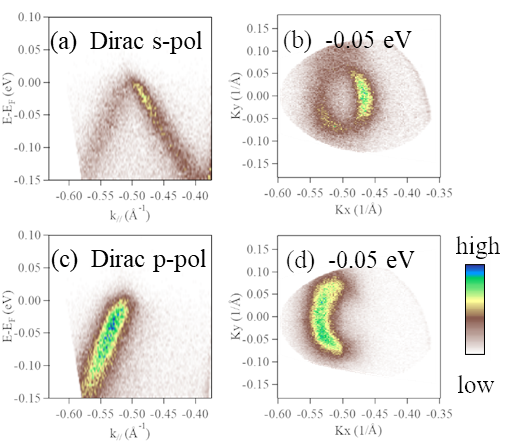}
	\caption{ARPES of Dirac cone states. (a) $E-k$ plot with s polarization. (b) The CEC measured with s polarization at $E-E_F = -0.05$ eV. (c) $E-k$ plot with p polarization. (d) The CEC measured with p polarization at $E-E_F = -0.05$ eV.}\label{fig5}
\end{figure}

With our 6.2 eV photon energy and manipulator rotation angle, we could reach the Dirac cone by changing the polar angle in $\rm{\Gamma M}$ configuration. Fig. \ref{fig5} shows the polarization-dependent data around one of the Dirac nodes. We could see that in p polarization only the left branch was observed while in s polarization both branches show up with different intensities, therefore the left should be dominated by $d_{z^2}$ orbital with some trace of $d_{x^2-y^2}$ orbital and the right branch should be attributed to $d_{x^2-y^2}$. This is consistent to what we observed in ref. \cite{bib1,bib5}.

\begin{figure}[htb]%
	\centering
	\includegraphics[width=0.6\textwidth]{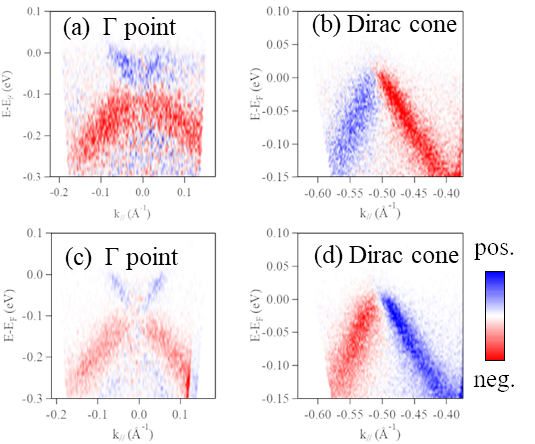}
	\caption{Comparison between difference data taken with (a) CL and CR polarization around $\rm{\Gamma}$ point. (b) CL and CR polarization around Dirac cone. (c) s and p polarization around $\rm{\Gamma}$ point. (d) s and p polarization around Dirac cone.}\label{fig6}
\end{figure}

Circular Dichroisms (CDs) of $\rm{BaNiS_2}$ around $\Gamma$ point and Dirac cone were investigated and plotted in Fig. \ref{fig6}(a)(b). The data were obtained in $\rm{\Gamma M}$ configuration. The CD we showed here were defined as:

\begin{equation}
I_{CD} = I_{CL}/ I_{max1} - I_{CR}/ I_{max2}
\end{equation}
where $I_{max1}$ and $I_{max2}$ are the maximum of $I_{CL}$ and $I_{CR}$, respectively. We can see that there were strong CD in B bands around $\rm{\Gamma}$ point and Dirac cone. Even though the formation mechanism of Dirac cone in $\rm{BaNiS_2}$ is different from topological insulators  \cite{bib5}, it shows a similar CD pattern compared to well-known topological insulators like $\rm{Bi_2Se_3}$ \cite{bib19,bib20}. In Fig. \ref{fig6}(c)(d) we plot the difference data between s and p polarization for comparison. It shows that around $\rm{\Gamma}$ point the linear and circular dichroism patterns have remarkable similarity, while around Dirac cone the linear and circular dichroism patterns are reversed, which strongly suggests that also the circular dichroism is related to the orbital nature of the bands. 

\section{Conclusions}\label{sec4}

We investigated the orbital character of $\rm{BaNiS_2}$ by using time-resolved ARPES with various light polarizations and for two crystal orientations with respect to the measurement geometry. By applying the selection rules, we determined the orbital characters of three bands around $\rm{\Gamma}$ and two branches of Dirac cones which are located in the $\rm{\Gamma M}$ direction. Our data indicate that the three bands A, B and C around $\rm{\Gamma}$ point consist of $d_{x^2-y^2}$, $d_{xy}$ and $d_{x^2-y^2}$ orbitals, respectively; where as the Dirac cones in $\rm{\Gamma M}$ direction have a $d_{z^2}$ and $d_{x^2-y^2}$ character. Our results demonstrate that dichroism in the time-resolved ARPES yield can be a valuable tool to investigate the orbital nature of filled and photoexcited electronic states in quantum materials.

\section*{Acknowledgements}

 M.M., L.P. and E.P. work was supported by "Investissement d'avenir Labex Palm" (Grant No. ANR-10-LABX-0039-PALM, by the Région Ile-de-France and in part by the France Berkeley Fund. J. Z. thanks the China Scholarship Council (CSC) for the financial support. 




\end{document}